\documentclass[onecolumn,showpacs,preprintnumbers,aps,prd,a4paper,superscriptaddress,nofootinbib,tightenlines,floats,floatfix ]{revtex4}
\usepackage{graphicx}
\usepackage{latexsym}
\usepackage{%amsmath,
amssymb}        % amssymb includes amsfonts

\newcommand\eqref[1]{(\ref{#1})}

\newcommand\ee{\end{equation}}
\newcommand\be{\begin{equation}}
\newcommand\eea{\end{eqnarray}}
\newcommand\bea{\begin{eqnarray}}

\newcommand\bfk{{\bf k}}

\newcommand\bfx{{\bf x}}

\newcommand{\vev}[1]{{\langle #1 \rangle }}

\begin{document}
%%%%%%%%%%%%%%%%%%

\title{Higher Order Contributions to the Primordial Non-Gaussianity}
\author{Ignacio~Zaballa}
\affiliation{Department of Physics, Lancaster University, Lancaster LA1 4YB, 
United Kingdom}
%\author{Yeinzon Rodr\'iguez}
%\affiliation{Physics Department, University of Lancaster, Lancaster LA1 4YB, 
%United
%Kingdom}
\author{Yeinzon~Rodr\'{\i}guez}
\affiliation{Escuela de F\'{\i}sica, Universidad Industrial de Santander, Ciudad Universitaria, Bucaramanga, Colombia}
\affiliation{Centro de Investigaciones, Universidad Antonio Nari$\tilde n$o, Cll 58A \# 37 - 94, Bogot\'a D.C., Colombia}
\author{David H.~Lyth}
\affiliation{Department of Physics, Lancaster University, Lancaster LA1 4YB, 
United Kingdom}
%\date{\today}
\pacs{98.80.-k \hfill JCAP 001P 0406, astro-ph/0603534}
%\preprint{astro-ph/0603534}

\begin{abstract}
In this paper we calculate additional contributions to that part of the non-Gaussianity of the primordial curvature perturbation $\zeta$, which come from the three-point correlator of the field perturbations. We estimate this contribution in the following models for the origin of $\zeta$: single-component inflation, multi-component chaotic inflation, a two-component ``hybrid'' inflationary model, and the curvaton scenario. In all of these models, the additional contributions to the primordial non-gaussianity considered here are too small to ever be detected.
\end{abstract}

\maketitle

%%%%%%%%%%%%%%%%%%%%%%%
\section{Introduction}
%%%%%%%%%%%%%%%%%%%%%%%

There is substantial interest in an observable deviation from a Gaussian distribution of the primordial curvature perturbation $\zeta$ \cite{Bartolo:2001cw,Acquaviva:2002ud,Maldacena:2002vr,Lyth:2005du,Lyth:2005fi,Boubekeur:2005fj,Rodriguez:2005th,Seery:2005gb,Rigopoulos:2005fi,Rigopoulos:2005ae,Lyth:2005qj,Barnaby:2006cq,Alabidi:2005qi,Allen:2005ye,Vernizzi:2006ve}. From observation, the nongaussian contribution of $\zeta$ is expected to be small \cite{Komatsu:2003fd}. Nevertheless the current bound still includes a wide range of models for the origin of $\zeta$, and therefore a future nongaussian detection or a tighter bound on the nongaussianity of $\zeta$, could rule out a considerable number of models.

The level of nongaussianiaty in the curvature perturbation can be parameterised by $f_{\rm NL}$, which specifies the three-point correlator of $\zeta$. Here we apply the $\delta N$ formalism \cite{Sasaki:1995aw,Lyth:2004gb} to calculate new contributions to $f_{\rm NL}$. While the relevant scales are outside the horizon, the curvature perturbation according to the $\delta N$ formalism,
is given by \cite{Lyth:2005fi}
\bea
\zeta(\bfx,t) &=& \delta N(\rho(t),\phi_i(\bfx)) \label{deln1} \\
&=& \sum_i N_i(t) \delta\phi_i(\bfx) + \frac12
\sum_{ij} N_{ij}(t) \delta\phi_i(\bfx) \delta\phi_j(\bfx)
+ \cdots \label{deln2}
\,,
\eea
where $N(\rho(t),\phi_i(\mathbf x))$ is the number of e-foldings from an initial flat slice, on which the field values are evaluated, to an uniform energy density slice at a time $t$ after $\zeta$ has settled down to the time-independent value. $N_i$ and $N_{ij}$ are the partial derivatives, $N_i=\partial N/\partial\phi^i$ and $N_{ij}=\partial^2 N/\partial\phi^i\partial\phi^j$, resulting from the Taylor expansion on the field perturbations. From now on, we consider only the first two terms of the expansion in Eq.(\ref{deln2})

In the case of Gaussian field perturbations $f_{NL}$ is given by \cite{Lyth:2005fi,Boubekeur:2005fj}
\be\label{eq:fnlgaussian}\label{fnl}
\frac{6}{5}f_{\rm NL}\simeq
\frac{\sum_{ij}N_{i}N_{j}N_{ij}}{\left(\sum_lN_{l}^2\right)^2}+\mathcal{P}_\zeta\frac{\sum_{ijk}N_{ij}N_{jk}N_{ki}}{\left(\sum_lN_{l}^2\right)^3}
\,.
\ee

Seery and Lidsey \cite{Seery:2005gb} have calculated the three-point correlator of the field perturbations. To leading order in the field perturbations, the three-point correlator is  
\be
\langle\zeta_{\mathbf k_1}\zeta_{\mathbf k_2}\zeta_{\mathbf k_3}\rangle
\subset\sum_{ijk}N_iN_jN_k\langle\delta\phi^i_{\mathbf k_1}\delta\phi^j_{\mathbf k_2}\delta\phi^k_{\mathbf k_3}\rangle \,,
\ee
which gives the following additional term
\be\label{eq:deltaf_nl}
\Delta f_{NL}=\frac{\sum_{ijk}N_iN_jN_k\;f^{ijk}\left(k_1,k_2,k_3\right)}
{\lbrack\sum_lN_l^2\rbrack^{3/2}{\cal P}_{\zeta}^{1/2}} \,.
\ee
The $f^{ijk}$'s are dimensionless functions proportional to the three-point correlator of the field perturbations. This term adds a contribution to $f_{\rm NL}$ too small to ever be observable, $\left|\Delta f_{\rm NL}\right|\leq0.044$ \cite{Lyth:2005qj}.  

%Taking the three point function calculated in Ref.\cite{Seery:2005gb},
%$f^{ijk}$ is given by
%\be
%f^{ijk}\left(k_1,k_2,k_3\right)=-\frac{5}{6}
%\frac{\sum_{\sigma'}\dot\phi_*^i\delta^{jk}{\cal M}_{123}\left(k_1,k_2,k_3\right)}{2\pi M_p^2\sum_lk_l^3}.
%\ee

%Here the sum $\sigma'$ is over all the permutations of the indices i, j and k, at the same time their respective momenta $k_1$, $k_2$ and $k_3$, and $\dot\phi_*^i$ is the time derivative of the field evaluated at horizon crossing during inflation. 

%When we calculate the two terms in Eq.(\ref{eq:deltaf_nl}), we can factorise the momenta dependent term, which has a maximum of for $k=k'=k''$ given by \cite{Lyth:2005qj}

%\be
%\left| \frac{\sum_{\sigma}\mathcal{M}_{123}\left(k,k',k''\right)}{\sum_lk_l^3} \right|_{\rm max}
%=\frac{11}{6}.
%\ee

%In the equation above, the sum over $\sigma$ denotes the sum over the permutations of the three momenta only. 

%The first term in (\ref{eq:deltaf_nl}) that we denote by $\Delta f_{NL}^{(1)}$, 
%$\Delta f_{NL}$ was calculated for a canonically normalized scalar field theory minimally coupled to gravity with an arbritary potential in \cite{Lyth:2005qj}. 

In this paper we calculate the next order contribution to $f_{NL}$. Contrary to the $\Delta f_{NL}$ case, we are unable to give an estimate for an arbitrary potential. Therefore we calculate its contribution to the nongaussianity of $\zeta$ in several specific cases: single field inflation, multicomponent chaotic inflation, the two-component inflationary model of Ref. \cite{Enqvist:2004bk}, and in the curvaton scenario. 

%We are unable to give a general estimate of the second term in Eq.(\ref{eq:deltaf_nl}), here $\Delta f_{\rm NL}^{(2)}$. Here we calculate its contribution to the nongaussianity of $\zeta$ in several specific cases: multicomponent (\textit{natural?})chaotic, the two-component inflationary model of Ref. \cite{Enqvist:2004bk}, and in the curvaton model. We will give only expressions of $\Delta f_{\rm NL}^{(2)}$ in the following sections for the case $k=k'=k''$, which maximizes the momenta dependence of the $\Delta f_{NL}$ function.

%%%%%%%%%%%%%%%%%%%%%%%%%%%%%%%%
\section{Higher order corrections to $f_{NL}$}
\label{f_NLcalc}
%%%%%%%%%%%%%%%%%%%%%%%%%%%%%%%%
The observed distribution of the curvature perturbation $\zeta$ in the CMB is almost Gaussian. The power spectrum
in the $\delta N$ formalism is then approximately given by
\be
\mathcal P_{\zeta}\simeq\left(\frac{H}{2\pi}\right)^2\sum_iN^2_i \,.
\ee 
The three-point correlator of $\zeta$, or its bispectrum
 defined by $\vev{\zeta_{\bfk_1}\zeta_{\bfk_2}\zeta_{\bfk_3}}
=(2\pi)^3B_{\mathcal{\zeta}}\;\delta^3(\mathbf{k}_1+\mathbf{k}_2
+\mathbf{k}_3)$, is the lowest order signature of  non-Gaussianity.
To measure a possible small deviation from gaussianity in $\zeta$ we introduce the $f_{NL}$ function, defined as \cite{Maldacena:2002vr}
\begin{equation}\label{eq:bispectrum}
B_{\mathcal{\zeta}}\left(k_1,k_2,k_3\right)=\frac{6}{5}f_{\rm
NL}\left[P_{\zeta}\left(k_1\right)P_{\zeta}\left(k_2\right)+\rm{cyclic}\right] \,,
\end{equation}
where $P_{\zeta}(k)=2\pi^2\mathcal{P}_\zeta/k^3$.

We consider the contribution of the three-point correlator of the field perturbations to the following term\footnote{Note that we have diagonalized the $N_{ij}$ in the field perturbations basis to simplify the calculation.
The connected part of the five point function is not written explicitely.}
\begin{widetext}
\be\label{eq:fivesum}
\langle\zeta_{\mathbf k_1}\zeta_{\mathbf k_2}\zeta_{\mathbf k_3}\rangle
\subset\frac{1}{4}\sum_{ijk}N_iN_{jj}N_{kk}\langle\delta\phi^i_{\mathbf k_1}\left(\delta\phi^j\right)^2_{\mathbf k_2}\left(\delta\phi^k\right)^2_{\mathbf k_3}\rangle\nonumber
+\left(k_1\leftrightarrow k_2\right)+\left(k_1\leftrightarrow k_3\right) \,,
\ee
\end{widetext}
where $k_1\leftrightarrow k_2$ and $k_1\leftrightarrow k_3$ represent the corresponding permutations, and $\left(\delta\phi\right)^2_{\mathbf k}$ denotes the convolution product of the field perturbations:
\be
\left(\delta\phi^2\right)_{\mathbf k}=\frac{1}{\left(2\pi\right)^3}
\int d^3\mathbf q\;\delta\phi_{\mathbf q}\delta\phi_{\mathbf k-\mathbf q} \,.
\ee 

The correlator in the field perturbations has the following generic form
\be\label{eq:fivepoint}
\langle\delta\phi^i_{\bf k_1}\delta\phi^j_{\bf q_2}\delta\phi^j_{\bf k_2-q_2}\delta\phi^k_{\bf q_3}\delta\phi^k_{\bf k_3-q_3}\rangle
=\langle\delta\phi^i_{\bf k_1}\delta\phi^j_{\bf q_2}\rangle
\langle\delta\phi^j_{\bf k_2-q_2}\delta\phi^k_{\bf q_3}\delta\phi^k_{\bf k_3-q_3}\rangle+{\rm Comb} \,,
%+\langle\delta\phi\delta\phi\delta\phi\delta\phi\delta\phi\rangle_C,
\ee 
where Comb refers to all possible combinations of the 5 elements in groups of two, which in total number $10$. In the equation above, the two point function is defined by
\be
\langle\delta\phi^i_{\bf k_1}\delta\phi^j_{\bf k_2}\rangle=\left(2\pi\right)^3\delta^{ij}\;P_{\delta\phi^i}\left(k_1\right)\delta^{(3)}\left(\mathbf k_1+\mathbf k_2\right) \,,
\ee
where we consider a scale invariant spectrum $\mathcal P_{\delta\phi^i}(k)=(k^3/2\pi^2)P_{\delta\phi^i}=(H/2\pi)^2$. On the other hand, the three-point function obtained from second order cosmological perturbation theory is \cite{Seery:2005gb}
\be\label{eq:threepoint}
\vev{\delta\phi^i_{\mathbf{k}_1}\delta\phi^j_{\mathbf{k}_2}
\delta\phi^k_{\mathbf{k}_3}}=(2\pi)^3\frac{\delta^{(3)}(\sum_i\mathbf{k}_i)}
{\prod_lk_l^3}\frac{4\pi^4}{M_P^2}
\left(\frac{H}{2\pi}\right)^4\sum_{\sigma'}\frac{\dot\phi_i}{2H}\;
\delta_{jk}\mathcal{M}_{123} \,,
\ee
with $\mathcal{M}_{123}$ given by
\begin{equation}\label{eq:Mfunction}
\mathcal{M}_{123}\equiv\mathcal{M}\left(k_1,k_2,k_3\right)=\frac{1}{2}\left[-3
\frac{k^2_2k_3^2}{k_t}-\frac{k^2_2k_3^2}{k_t^2}\left(k_1+2k_3\right)+
\frac{1}{2}k_1^3-k_1k_2^2\right] \,,
\end{equation}
where $k_t$ denotes the sum of the modulus of the three momenta $k_t=k_1+k_2+k_3$. In Eq. (\ref{eq:threepoint}) the sum $\sigma'$ is over all the permutations of the indices $i$, $j$ and $k$, and at the same time over their respective momenta $k_1$, $k_2$ and $k_3$. $M_P$ is the reduced Plank mass: $M_P=2.436\times10^{18}$ GeV. 

\begin{figure*}
\includegraphics[angle=0,width=0.5\textwidth]{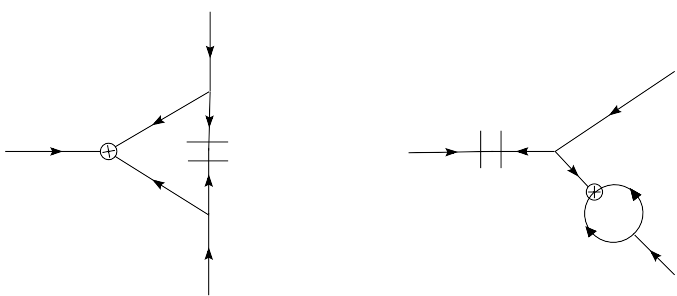}
\caption{\label{fig:1}
Here we show the two possible diagrams. There are three external lines corresponding to the three momenta $k_1$, $k_2$ and $k_3$. The parallel lines represent the two-point correlator, whereas the crossed circle indicates the three-point correlator. The diagrams have been drawn using JaxoDraw \cite{Binosi:2003yf}.
}
\end{figure*}

Of the $10$ combinations in Eq. (\ref{eq:fivepoint}) only $8$ are non-vasnishing, since neither $\mathbf k_2$ nor $\mathbf k_3$ can be zero. The $8$ terms can be arranged into two groups of $4$ terms, each group containing terms that give the same contribution. These can be represented schematically as two different diagrams, shown in Fig.(\ref{fig:1}). The other two contributions in Eq. (\ref{eq:fivesum}) arising from permuting the $k_1$ momentum with $k_2$ and $k_3$, result in the same two diagrams. Similar diagrams are given in Ref. \cite{Crocce:2005xy}.

In calculating the contribution to the three-point function of $\zeta$ resulting from the first diagram, we find the following integral over the internal momentum:
\be
I_{123}=\int_{L^{-1}}\frac{d^3\mathbf q}{q^3}\;
\frac{\mathcal{M}\left(k_1,\vert\mathbf k_2+\mathbf q\vert,\vert\mathbf k_3-\mathbf q\vert\right)}
{k_1^3\vert\mathbf k_2+\mathbf q\vert^3\vert\mathbf k_3-\mathbf q\vert^3} \,, 
\ee 
and similar integrals corresponding to the permutations in the argument of the $\mathcal{M}_{123}$ function. The momenta dependance of the integrand can be read off from the location of the three-point correlator in Fig.(\ref{fig:1}). In the equation above $L^{-1}$ indicates that the integrand is set equal to zero in a sphere of radius $L^{-1}$ around each singularity. The integral then is approximately given at each singularity by
\be\label{eq:integrals}
I_{123}=4\pi\;{\rm ln}KL\;\left(\frac{\mathcal{M}\left(k_1,\vert\mathbf k_2+\mathbf q\vert,\vert\mathbf k_3-\mathbf q\vert\right)}
{k_1^3\vert\mathbf k_2+\mathbf q\vert^3\vert\mathbf k_3-\mathbf q\vert^3}\right)_{\mathbf q=0, -\mathbf k_2\;{\rm or}\;\mathbf k_3} \,,
\ee
where the function in brackets is evaluated at the singularity, and $K={\rm min}\{k_i\}$. The logarithm in the expression above can be taken to be of order 1 \cite{Boubekeur:2005fj}. It can be shown that each permutation of the $\mathcal{M}_{123}$ function only diverges at $q=0$, and only at one of the other two singularities if at any, since the integrand becomes $0$ in at least one of them. Therefore we are able to give a momenta dependent result for each integral.   

The total contribution to the the $f_{NL}$ function from the first diagram, using the identity $\sum_iN_i\dot\phi^i=-H$ as in \cite{Vernizzi:2006ve},  gives
\be\label{eq:deltafnl1}
\frac{6}{5}\Delta f^{(1)}_{NL}=\frac{1}{M^2_p}\left(\frac{H}{2\pi}\right)^4\frac{\left[2\sum_{i}N_iN_{ii}^2\left(\frac{\dot\phi^i}{2H}\right)-\frac{1}{2}\sum_jN_{jj}^2\right]
\frac{\sum_{\sigma}\mathcal{M}_{123}}{\sum_lk_l^3}-4\sum_{i}N_iN_{ii}^2\left(\frac{\dot\phi^i}{2H}\right)}{\sum_kN_k^2}.
\ee 

The momenta dependent term has a maximum absolute value for $k\equiv k_1=k_2=k_3$ given by \cite{Lyth:2005qj}
\be
\frac{\sum_{\sigma}\mathcal{M}\left(k,k,k\right)}{\sum_lk_l^3}
=-\frac{11}{6} \,,
\ee
where the sum over $\sigma$ denotes the sum over the permutations of the three momenta. In the following sections, we calculate the value of $\Delta f^{(1)}_{NL}$ for the maximum of the momenta dependent term. 

The integrals for the second diagram are evaluated in a similar way to the integrals of the first diagram, but this time there are more cancellations at the singularities. The total contribution to $f_{NL}$ function from the second diagram is
\be\label{eq:deltafnl2}
\frac{6}{5}\Delta f^{(2)}_{NL}=-\frac{4}{M^2_p}\left(\frac{H}{2\pi}\right)^4\left[\frac{\sum_{i}N_iN_{ii}^2\left(\frac{\dot\phi^i}{2H}\right)}{\sum_kN_k^2}\right] \,.
\ee 

In the following sections, we evaluate these higher order contributions in Eqs. (\ref{eq:deltafnl1}) and (\ref{eq:deltafnl2}) for several specific cases.

%%%%%%%%%%%%%%%%%%%%%%%%%%%%%%%%
\section{Single-component Inflation}
\label{Single}
%%%%%%%%%%%%%%%%%%%%%%%%%%%%%%%%

In single field inflation the number of e-foldings $N(t,t_0)$ is given by
\be
N\left(t,t_0\right)\equiv\int^t_{t_0} Hdt\simeq\frac{1}{M^2_p}\int^{\phi_0}_\phi\frac{V}{V_\phi}\;d\phi \,.
\ee

In the $\delta N$ formalism we evaluate the partial derivatives of $N$ respect of the field values on the initial flat slice. Then
\bea
N_\phi&=&\frac{1}{\sqrt{2\epsilon}M_p} \,,\\
N_{\phi\phi}&=&\frac{1}{M^2_p}\left(1-\frac{\eta}{2\epsilon}\right) \,,
\eea
where $\epsilon=M^2_p/2\;(V_\phi/V)^2$ and $\eta=M^2_p(V_{\phi\phi}/V)$ are the slow-roll parameters.

%Taking the first term in Eq.(\ref{fnl}) and the correction from the three-point correlator of the fields, we recover Maldacena's result for single field inflation 

Defining $\Delta f_{NL}^{(1+2)}\equiv\Delta f_{NL}^{(1)}+\Delta f_{NL}^{(2)}$, we have the following contribution to $f_{NL}$:
\be
\frac{6}{5}\Delta f_{NL}^{(1+2)}=54\;\mathcal P_{\zeta}^2\left(\epsilon^3-\eta\epsilon^2+\frac{\eta^2\epsilon}{4}\right) \,.
\ee

This is much less than the leading order contribution calculated in \cite{Seery:2005gb}.  
%%%%%%%%%%%%%%%%%%%%%%%%%%%%%%%%
\section{Multi-component Chaotic Inflation}
\label{multi_com}
%%%%%%%%%%%%%%%%%%%%%%%%%%%%%%%%

In this section we consider a potential with the following form \cite{Dimopoulos:2005ac,Alabidi:2005qi}
\be
V=\frac{1}{2}\sum_{i=1}^nm_i^2\phi_i^2 \,.
\ee

This potential represents multi-component chaotic inflation. Under slow-roll conditions the masses of the fields are similar \cite{Polarski:1992dq}, and then the number of e-foldings $N$ is given by $N\left(\phi_i,\rho\right)=\frac{1}{4}\sum_{i=1}^n(\phi_i^2/M_P^2)$ \cite{Lyth:1998xn}. The derivatives of $N$ are
\bea
N_i=\phi_i/2M_p^2 \,, \\
N_{ij}=\delta_{ij}/2M_p^2 \,.
\eea

Then the contribution in Eq. (\ref{eq:fnlgaussian}) to $f_{NL}$ is much less than one \cite{Alabidi:2005qi}; $(6/5)f_{NL}=1/N+O(1/N^3)$, and therefore it is not observable. The higher order contributions in Eqs. (\ref{eq:deltafnl1}) and (\ref{eq:deltafnl2}) give
%The contribution of $\Delta f_{NL}^{(1)}$ is of the same order of magnitude as $f_{NL}$ 
%\be
%\vert\frac{\Delta f_{NL}^{(1)}}{f_{NL}}\vert=\frac{11}{12}.
%\ee
\be
\frac{6}{5}\Delta f_{NL}^{(1+2)}=\left(\frac{r}{16}\right)^3\mathcal P^2_{\zeta}
\left[\frac{11}{6}\left(n+2\right)+8\right] \,,
\ee
where $r$ is the tensor to scalar ratio $r=8(H/2\pi)^2/M_p^2{\cal P}_{\zeta}$, which in this case is given by $r=8/N$. This contribution is much smaller than the leading order one in Eq. (\ref{eq:fnlgaussian}), and therefore negligible.  

%\be
%\frac{\Delta f_{NL}^{(1+2)}}{f_{NL}}\sim\frac{{\cal P}^2_{\zeta}}{N^2}
%\;.
%\ee

%and comparable to the second term in equation (\ref{eq:fnlgaussian}) giving a contribution $O(1/N^3)$.
 
%%%%%%%%%%%%%%%%%%%%%%%%%%%%%%%%%%%%%%%%%%%%%%%%%%%%%
\section{Two-component model}
\label{two_com}
%%%%%%%%%%%%%%%%%%%%%%%%%%%%%%%%%%%%%%%%%%%%%%%%%%%%%

Here we consider a two-component model given by the following potential \cite{Lyth:2005du,Enqvist:2004bk}:
\be
V\left(N,\phi,\sigma\right)=V_0\left[1+\frac{1}{2}\eta_{\phi}\frac{\phi^2\left(N\right)}{M_p^2}+
\frac{1}{2}\eta_{\sigma}\frac{\sigma^2\left(N\right)}{M_p^2}\right] \,,
\ee
where $\eta_{\phi}$ and $\eta_{\sigma}$ are the $\eta$ slow-roll parameters. $\phi(N)$ and $\sigma(N)$ are given in terms of the field values just after horizon exit, $\phi$ and $\sigma$, by $\phi(N)=\phi\;\rm{exp}(-N\eta_{\phi})$ and $\sigma(N)=\phi\; \rm{exp}(-N\eta_{\sigma})$. Here we consider the case $\sigma=0$ of Ref. \cite{Enqvist:2004bk}.
The derivatives of N are \cite{Lyth:2005fi,Malik:2005cy}
\bea
N_{\phi}&=&\left(1/\eta_{\phi}\phi\right),\;N_{\sigma}=0 \,,\\
N_{\phi\phi}&=&-\eta_{\phi}(1/\eta_{\phi}\phi)^2 \,,\\ N_{\sigma\sigma}&=&\eta_{\sigma}(1/\eta_{\phi}\phi)^2\;\rm{e}^{2N(\eta_{\phi}-\eta_{\sigma})} \,. 
\eea
%\textit{Shouldnt the exponetial have a negative sign? Otherwise the contribution I get below to $f_{NL}$ in the case considered in Ref.\cite{Lyth:2005fi}, $\eta_{\phi}\geq0.26$, $\eta_{\sigma}= \eta_{\phi}/2$, and $\zeta_{\sigma}=10^{-2}\zeta$, is too large. Actually, note that in the expression below i do not need $\zeta_{\sigma}=10^{-2}\zeta$.}

%$N_{\phi}=1/\eta_{\phi}\phi$, $N_{\sigma}=0$, $N_{\phi\phi}=-\eta_{\phi}(1/\eta_{\phi}\phi)^2$, and $N_{\sigma\sigma}=\eta_{\sigma}(1/\eta_{\phi}\phi)^2\;\rm{exp}[2N(\eta_{\phi}-\eta_{\sigma})]$ \cite{Lyth:2005fi}.

The $f_{NL}$ function is dominated by the term in Eq. (\ref{eq:fnlgaussian}) proportional to $N_{\sigma\sigma}$, and it is approximately given by
\be
\frac{6}{5}f_{\rm NL}=\left(\frac{r}{16}\right)^3\mathcal{P}_\zeta
\left[\left(\frac{\eta_{\sigma}}{\epsilon}\right)^3{\rm e}^{6N\left(\eta_{\phi}-\eta_{\sigma}\right)}\right] \,,
\ee 
where $\epsilon$ is the slow roll parameter across the $\sigma=0$ trajectory, $\epsilon=(\eta_{\phi}\phi/\sqrt{2}M_p)^2$, and $r$ is the tensor to scalar ratio defined earlier. Similarly, $\Delta f_{NL}^{(1+2)}$ is dominated by the term proportional to $N_{\sigma\sigma}$, and therefore the main contribution comes from $\Delta f_{NL}^{(1)}$:
%The $\Delta f_{NL}^{(1)}$ function gives
%\be
%\vert\Delta f_{NL}^{(1)}\vert=0.12\;re^{-2N\eta_{\phi}} \,,
%\ee
\be
\frac{6}{5}\Delta f_{NL}^{(1)}=\frac{11}{6}\left(\frac{r}{16}\right)^3\mathcal{P}^2_\zeta 
\left[\left(\frac{\eta_{\sigma}}{\epsilon}\right)^2{\rm e}^{4N\left(\eta_{\phi}-\eta_{\sigma}\right)}\right] \,.
%\frac{\sum_{\sigma}\mathcal{M}_{123}}{\sum_lk_l^3}
\ee

This contribution is much smaller than $f_{NL}$;
\be
\frac{\Delta f_{NL}^{(1)}}{f_{\rm NL}}\simeq\mathcal{P}_\zeta\left(\frac{\eta_{\sigma}}{\epsilon}\right)^{-1}
{\rm e}^{-2N\left(\eta_{\phi}-\eta_{\sigma}\right)} \,. 
\ee
     
%%%%%%%%%%%%%%%%%%%%%%%%%%%%%%%%%%%%%%%%%%%%%%%%%%%%%%%%%%%%%%%%%%%%%%%%%
\section{Curvaton Scenario}
\label{curvaton_mod}
%%%%%%%%%%%%%%%%%%%%%%%%%%%%%%%%%%%%%%%%%%%%%%%%%%%%%%%%%%%%%%%%%%%%%%%%%
%
In the curvaton scenario, the curvature perturbation $\zeta$ is generated during the oscillations of the curvaton field $\sigma$ after the end of inflation \cite{Lyth:2001nq,Moroi:2001ct,Lyth:2003uw}. The curvaton starts oscillating around the minimum of a quadratic potential, $V(t,\mathbf{x})=m^2\sigma^2(t,\mathbf{x})/2$, in an initially radiation dominated epoch with amplitude $g\equiv g(\sigma_*)$ until it eventually decays. Here $\sigma_*$ is the value of the curvaton field just after horizon exit during inflation. Defining $N$ as the number of e-foldings from the beginning of the oscillations to the curvaton decay, its derivatives are given by \cite{Lyth:2005fi} 
\bea
N_{\sigma_*}&=&\frac{2\Omega_\sigma g'}{3g} \,,\\
N_{\sigma_*\sigma_*}&=&\frac{2\Omega_\sigma}{9g^2}\left[2\Omega_\sigma\left(2+\Omega_\sigma\right)g'^2-3\left(g'^{2}+gg''\right)\right] \,,
\eea
where $g'$ and $g''$ denote the first and second derivatives with respect to $\sigma_*$, and $\Omega_\sigma$ is the final fraction of the curvaton energy density before it decays. Then \cite{Lyth:2005fi,Lyth:2005du}
\be
f_{NL}=-\frac{5}{3}-\frac{5}{6}\Omega_\sigma+\frac{5}{4}\Omega_\sigma\left(1+\frac{gg''}{g'^2}\right) \,.
\ee

%\textit{Change notation for the final fraction of the curvaton energy density?} 
%$N_{\sigma_*}=2sg'/3g$ and $N_{\sigma_*\sigma_*}=(2s/9g^2)[2s(2+s)g'^2-3(g'2+gg'')$ 

%Defining $\epsilon_{\sigma}=M_p^2(V_{\sigma}/\sqrt{2}V)^2$, where $V_{\sigma}=\partial V/\partial\sigma$, the first term in eq.(\ref{eq:deltaf_nl}) gives $\Delta f_{NL}^{(1)}=-(11/24)\sqrt{\epsilon_{\sigma}r}$. If the fraction of energy density of the curvaton dominates, that is $\Omega_{\sigma}\simeq1$, then ($f_{NL}\simeq+5/4$ and therefore)\\
%\textit{Note that r is the gravitational waves fraction}

%\be
%\vert\frac{\Delta f_{NL}^{(1)}}{f_{NL}}\vert\simeq0.6\sqrt{\epsilon_{\sigma}r}.
%\ee
  
%If $\Omega_{\sigma}\ll1$, then $f_{NL}$ is dominated by the negative term and it can be very large \cite{Lyth:2005du}. Then the $\Delta f_{NL}^{(1)}$ function is again negligible.

Defining $\epsilon_{\sigma}=M_p^2(V_{\sigma}/\sqrt{2}V)^2$, where $V_{\sigma}=\partial V/\partial\sigma$, the relative contribution of $\Delta f_{NL}^{(1+2)}$ to $f_{NL}$ is
\be
\left(\frac{\Delta f_{NL}^{(1+2)}}{f_{NL}}\right)\simeq r\;{\cal P}^2_{\zeta}f_{\rm NL} \,,
\ee
being too small to be detected because of the current observational constraint, $f_{NL}<121$ \cite{Komatsu:2003fd}.
 
%%%%%%%%%%%%%%%%%%%%%%%%%%%%%%%%%%%%
\section{Conclusion}
\label{disc_sec}
%%%%%%%%%%%%%%%%%%%%%%%%%%%%%%%%%%%%

We have calculated a higher-order  contribution 
of the three-point correlator of the field perturbations
to the three-point correlator of the curvature perturbation, which is the next to leading order in the field perturbations. Even though intuition suggests that it should be smaller than the 
leading-order contribution, we have not been able to give a general proof
that this is so. Instead, we have shown that it is negligible for the 
cases that have been considered so far in the literature. Keeping only the first two terms of Eq.(\ref{deln2}), there is one higher order contribution, and more contributions will be generated by higher order terms in Eq.(\ref{deln2}). All of these contributions can be calculated in a similar way to the calculation shown here. 

%%%%%%%%%%%%%%%%
\acknowledgments
%%%%%%%%%%%%%%%%
IZ is partially supported by Lancaster University Physics Department.
YR is supported by COLCIENCIAS (COLOMBIA).
DHL is supported by PPARC grants PPA/G/S/2002/00098, PPA/G/O/2002/00469, 
PPA/Y/S/2002/00272, PPA/V/S/2003/00104 and EU grant MRTN-CT-2004-503369.

%%%%%%%%%%%%%%%%%%%%%%%%%%%%%%%%%%%%%%%%%%%%%%%%%%%%%%%%%%%%%%%%%%%%

%%%%%%%%%%%%%%%%%%%%%
%%%%%%%%%%%%%%%%%%%%%%%%%%%%%%%%%%%%%%%%%%%%%%%%%
\end{document}